\documentstyle[psfig]{mn}

\newif\ifAMStwofonts
\AMStwofontstrue

\newcommand{\be}{\begin{equation}}    
\newcommand{\ee}{\end{equation}}
\newcommand{\beq}{\begin{eqnarray}}
\newcommand{\eeq}{\end{eqnarray}}
\def\msun{M_\odot}
\def\op{ $ }
\def\cl{$ }
\def\nn{\nonumber}

\ifoldfss
  \ifCUPmtlplainloaded \else
    \NewTextAlphabet{textbfit} {cmbxti10} {}
    \NewTextAlphabet{textbfss} {cmssbx10} {}
    \NewMathAlphabet{mathbfit} {cmbxti10} {} 
    \NewMathAlphabet{mathbfss} {cmssbx10} {} 
  \fi
  \ifAMStwofonts
    \ifCUPmtlplainloaded \else
      \NewSymbolFont{upmath} {eurm10}
      \NewSymbolFont{AMSa} {msam10}
      \NewMathSymbol{\upi}     {0}{upmath}{19}
      \NewMathSymbol{\umu}     {0}{upmath}{16}
      \NewMathSymbol{\upartial}{0}{upmath}{40}
      \NewMathSymbol{\leqslant}{3}{AMSa}{36}
      \NewMathSymbol{\geqslant}{3}{AMSa}{3E}

      \let\leq=\leqslant 
       
    \fi
  \fi
\fi 

\ifnfssone
  \newmathalphabet{\mathit}
  \addtoversion{normal}{\mathit}{cmr}{m}{it}
  \addtoversion{bold}{\mathit}{cmr}{bx}{it}
  \newmathalphabet{\mathbfit} 
  \addtoversion{normal}{\mathbfit}{cmr}{bx}{it}
  \addtoversion{bold}{\mathbfit}{cmr}{bx}{it}
  \newmathalphabet{\mathbfss} 
  \addtoversion{normal}{\mathbfss}{cmss}{bx}{n}
  \addtoversion{bold}{\mathbfss}{cmss}{bx}{n}
  \ifAMStwofonts
    \ifCUPmtlplainloaded \else
      \UseAMStwoboldmath
      \makeatletter
      \new@mathgroup\upmath@group
      \define@mathgroup\mv@normal\upmath@group{eur}{m}{n}
      \define@mathgroup\mv@bold\upmath@group{eur}{b}{n}
      \edef\UPM{\hexnumber\upmath@group}
      \new@mathgroup\amsa@group
      \define@mathgroup\mv@normal\amsa@group{msa}{m}{n}
      \define@mathgroup\mv@bold\amsa@group{msa}{m}{n}
      \edef\AMSa{\hexnumber\amsa@group}
      \makeatother
      \mathchardef\upi="0\UPM19
      \mathchardef\umu="0\UPM16
      \mathchardef\upartial="0\UPM40
      \mathchardef\leqslant="3\AMSa36
      \mathchardef\geqslant="3\AMSa3E

      \let\leq=\leqslant 

    \fi
  \fi
\fi 

\ifnfsstwo
  \DeclareMathAlphabet{\mathbfit}{OT1}{cmr}{bx}{it}
  \SetMathAlphabet\mathbfit{bold}{OT1}{cmr}{bx}{it}
  \DeclareMathAlphabet{\mathbfss}{OT1}{cmss}{bx}{n}
  \SetMathAlphabet\mathbfss{bold}{OT1}{cmss}{bx}{n}
  \ifAMStwofonts
    \ifCUPmtlplainloaded \else
      \DeclareSymbolFont{UPM}{U}{eur}{m}{n}
      \SetSymbolFont{UPM}{bold}{U}{eur}{b}{n}
      \DeclareSymbolFont{AMSa}{U}{msa}{m}{n}
      \DeclareMathSymbol{\upi}{0}{UPM}{"19}
      \DeclareMathSymbol{\umu}{0}{UPM}{"16}
      \DeclareMathSymbol{\upartial}{0}{UPM}{"40}
      \DeclareMathSymbol{\leqslant}{3}{AMSa}{"36}
      \DeclareMathSymbol{\geqslant}{3}{AMSa}{"3E}

      \let\leq=\leqslant 

    \fi
  \fi
\fi 

\ifCUPmtlplainloaded \else
  \ifAMStwofonts \else 
    \def\upi{\pi}
    \def\umu{\mu}
    \def\upartial{\partial}
  \fi
\fi

\title[Gravitational-wave background from young, rapidly rotating neutron 
stars]
{Stochastic background of gravitational waves generated by
a cosmological population of young, rapidly rotating 
neutron stars}
\author[V. Ferrari, S. Matarrese and R. Schneider]
       {Valeria Ferrari$^1$, Sabino Matarrese$^2$ and Raffaella Schneider$^1$\\
        $^{1}$Dipartimento di Fisica ``G. Marconi",
Universit\'a degli Studi di Roma, ``La Sapienza"
and Sezione INFN  ROMA1,\\ p.le A.  Moro
5, 00185 Roma, Italy\\
        $^2$Dipartimento di Fisica ``Galileo Galilei ",
Universit\'a degli Studi di Padova
and Sezione INFN  PADOVA,\\ via Marzolo 8, 35131 Padova, Italy}

\date{June  1998}

\pagerange{\pageref{firstpage}--\pageref{lastpage}}
\pubyear{1998}
\begin{document}

\maketitle
\label{firstpage}

\begin{abstract}
We estimate the spectral properties of the
stochastic background of gravitational radiation emitted
by a cosmological population of hot, young, rapidly rotating neutron stars. 
Their formation rate as a function of redshift is deduced from
an observation-based determination of the star formation history in the
Universe, and the gravitational energy  is  assumed to be radiated
during the spin-down phase associated to the newly discovered r-mode 
instability. 
We calculate the overall signal produced by the ensemble of such neutron stars, 
assuming various cosmological backgrounds. 
We find that the spectral strain amplitude has a maximum 
$\approx (2-4)\times 10^{-26} \,\, \mbox{Hz}^{-1/2}$ , at frequencies 
$\approx (30-60)$ Hz, while the corresponding closure density, 
$h^2 \, \Omega_{GW}$, has a maximum amplitude plateau 
of $\approx (2.2-3.3) \times 10^{-8}$ in the frequency range $(500-1700)$
Hz.  We compare our results with a preliminary analysis
done by Owen et al. (1998), and discuss the  detectability of 
this background.

\end{abstract}


\section{Introduction}

In the last two years, a series of investigations on the perturbations of 
rotating relativistic stars have revealed the existence 
of a class of r-modes that are unstable due to the emission of 
gravitational radiation (Andersson 1998; Friedman \& Morsink 1998;  
Lindblom, Owen \& Morsink 1998). 
This gravitationally driven instability is of
considerable importance because it has a number of interesting astrophysical
implications. For instance, it determines a spin down of 
newly born neutron stars to  periods closer to the initial periods inferred
for ms pulsars. Moreover, as the star spins down an energy 
equivalent to $\approx 1\%$ of a solar mass
is emitted in gravitational waves making the process a very interesting source
for gravitational-wave detection (Owen et al. 1998;
Andersson, Kokkotas \& Schutz 1998).    

In this paper we study the stochastic background of 
gravitational waves contributed by a cosmological population of hot, young, 
rapidly rotating neutron stars with a small initial excitation in the $l=2$ 
r-mode. 
A rough estimate of the spectrum of this background radiation has recently
been given by Owen et al. (1998), assuming a 
comoving number density of neutron star births which is constant in the range 
$0<z<4$ and  zero at earlier times.

Here we present a more complete analysis, which derives the 
neutron star birth rate as a function of redshift 
from an observation-based determination of the star formation density 
evolution (Madau, Pozzetti \& Dickinson 1997), 
within three different cosmological background models.
Thus, the present investigation relies on a 
source rate which correctly accounts for the observed strong luminosity
evolution of star forming galaxies between $z = 0$ and $z=1-2$ (Lilly et al. 
1998; Connolly et al. 1997). With a similar approach, the gravitational
background emitted by an ensemble of stellar mass black holes has been
recently computed by Ferrari, Matarrese \& Schneider (1998). 

The plan of the paper is as follows. In Section 2 we calculate the universal 
rate of neutron star formation; in Section 3 we briefly introduce the 
adopted model of gravitational-wave emission by an
unstable  rapidly rotating neutron 
star, closely following the recent analysis by Owen et al. (1998);
in Section 4 we estimate the overall stochastic background produced by the 
ensemble of such sources throughout the Universe and calculate the total
duty cycle of the process; 
Section 5 contains a brief analysis of the detectability of such 
a background, while conclusions are drawn in Section 6. 

\section{The Rate of Neutron Star Formation in the Universe}

Type II and Type Ibc supernovae are fuelled by the gravitational contraction
of their core (e.g. Ruiz-Lapuente 1997). Numerical studies have shown that 
single stars with masses between $8$ and $20 \msun$ form 
iron cores with masses near the Chandrasekhar limit (e.g. Timmes, Woosley \& 
Weaver 1996). Thus, there is universal agreement that
these stars leave neutron star remnants. The fate of stars with masses 
greater than $20 \msun$  is less certain and depends on the iron core
masses formed and on the effect of fallback during the explosion
(e.g. Woosley \& Weaver 1995; Woosley \& Timmes 1996). 

For the calculations to follow, we assume that stars with masses between
$8$ and $25 \msun$ give origin to neutron stars.
However,   we will also  investigate the
effect of an upper mass cutoff of $\sim 20 \msun$.

The number of neutron stars formed per unit time within the comoving volume
out to redshift $z$ is given by (Ferrari, Matarrese \& Schneider 1998)
\be
R_{NS}(z)=\int_{0}^{z} \! \!\dot{\rho}_*(z') \,\, \frac{dV}{dz'} \,\,dz' 
\int_{8 \msun}^{25 \msun} \!\! \Phi(M) \,\,dM.
\label{rate}
\ee
Here $\dot{\rho}_*(z)$ is the star formation rate density (mass of gas that 
goes into stars per unit time and unit comoving volume),
$\Phi(M)$ is the initial mass function (IMF), and $dV/dz$ is the comoving
volume element. 

Following Madau, Pozzetti \& Dickinson (1997), we model the star formation
evolution from the recent collection of UV-optical observations of star
forming galaxies out to redshifts $\sim 4-5$, 
by assuming a Salpeter IMF,
\be
\Phi(M) \propto M^{-(1+x)} \;, \qquad \mbox{with} \qquad  x=1.35 \;,
\ee
normalized through the relation
\be
\int_{0.1 \msun}^{125 \msun} \!\! M\,\,\Phi(M)\,\,dM = 1.
\ee  
The star formation history is plotted in Figure 1 for a flat cosmology
with vanishing cosmological constant and $h=0.5$ (here $h$ is the Hubble 
constant, $H_0$, in units of $100$  km s$^{-1}$ Mpc$^{-1}$). 
The star formation rate rises sharply from its local value
to a peak at $z \sim 1.5$ to fall again out to $z\sim 4$. This behaviour 
indicates that the bulk of the stellar population was assembled in the 
redshift range 
$1\!\stackrel{\scriptscriptstyle <}{\scriptscriptstyle \sim}\!z\!\stackrel
{\scriptscriptstyle<}{\scriptscriptstyle \sim}\!2$.
This is consistent with the indications of QSO absorption lines and 
metallic clouds observations (Lanzetta, Yahil \& Fernandez-Soto 1996; Pei \& 
Fall 1995) and with the predictions of semi-analytic models for a broad class
of hierarchical clustering cosmologies (Baugh et al. 1998). 
A thorough review of the current state of observations of the star formation
history of field galaxies is given in Ellis (1997).  
\begin{figure}
\begin{center}
\leavevmode
\centerline{\psfig{figure=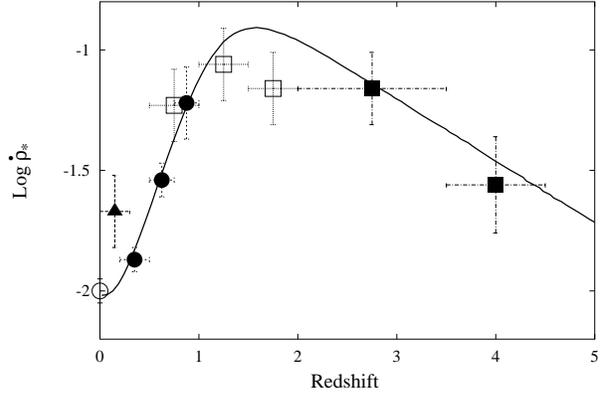,angle=270,width=8cm}}
\caption{Evolution of the SFR density
($\msun \mbox{yr}^{-1} \mbox{Mpc}^{-3}$) for $\Omega_0=1$,
($\Lambda=0$), $h = 0.5$ and a Salpeter IMF. The data points are 
taken from Gallego et al. (1995) ({\it empty dot}), Treyer et al. (1998) 
({\it filled triangle}), Lilly et al. (1996) ({\it filled dots}), 
Connolly et al. (1997) ({\it empty squares}) and the HDF (Madau et al. 1996) 
({\it filled squares}). The solid 
line represents the SFR density evolution assumed in the present analysis.
The fit has been kindly provided by P. Madau.}
\end{center}
\label{sfr}
\end{figure}
A summary of the relevant aspects for the present analysis
is given in Ferrari, Matarrese \& Schneider (1998).   

The predictions for the rate of core-collapse supernovae are in good 
agreement with the observed local values (Cappellaro et al. 1997).
At $z>1$, the detection of Type II events must await future 
experiments, such as the Next Generation Space Telescope (Madau, Della Valle
\& Panagia 1998).

Here we shall only stress that one major source of uncertainty in the
interpretation of the data points in Figure 1 is the effect
of dust extinction. Even a small amount of dust in primeval galaxies can
significantly attenuate the UV luminosity and re-radiate it in the far IR
thus leading to a potentially serious underestimate of the star formation
activity. Dust correction factors are still very uncertain mainly because of 
the unknown shape of the dust UV-extinction law (Pettini et al. 1997).
The model for the global SFR evolution presented in Figure 1
assumes an extinction law similar to that which applies to stars in the
Small Magellanic Cloud (SMC) and an amount of dust which results
in an upwards correction of the comoving UV luminosity by a factor $1.4$ 
at $2800$\AA\enspace and $2.1$ at $1500$\AA\enspace (Madau, Pozzetti \& 
Dickinson 1997). 
Recent near infrared observations of a
galaxy at $z=4.92$ has given evidence that absorption by dust is important up 
to redshifts of $\sim 5$ (Soifer et al. 1998). Similarly, analysis of the
Infrared Space Observatory (ISO) data have shown significant departure 
from the UV-optically derived SFR evolution at $z>2$ (Rowan-Robinson et al. 
1997).
 
This may seem a serious caveat for the present investigation. However,
this is not necessarily the case because the extra-galactic astrophysical 
gravitational-wave backgrounds prove to be mainly contributed by 
low-to-intermediate redshift sources. High redshift sources play a
negligible role because the energy flux 
emitted by each source decreases as the inverse of the squared luminosity
distance.
On the other hand, if an upwards correction of the UV-optically derived SFR 
density is required at $z \approx 1.5$, as suggested by a recent detection of 
the infrared background by the Diffuse Infrared Background Experiment (DIRBE) 
and the Far Infrared Absolute Spectrometer (FIRAS) on board of the 
Cosmic Background Explorer (COBE) satellite (Dwek et al. 1998), this would
result in an upwards correction of roughly the same amplitude on our 
gravitational-wave background predictions.  

We have evaluated equation (\ref{rate}) for three sensible 
cosmological models (e.g. Baugh et al. 1998), namely 
a flat geometry with vanishing cosmological constant, an open background
and a flat low-density model (see Table 1).
\begin{table}
\centering
\caption{Cosmological parameters adopted in the three backgrounds
considered.} \begin{tabular}{@{}llllll@{}}
\multicolumn{3}{c}{Model}&$\Omega_{M}$ & $\Omega_{\Lambda}$ & $h$  \\
 & & &   &   &    \\   
 &A& & 1 & 0 & 0.5 \\
 &B& & 0.3 & 0.7 & 0.6 \\
 &C& &0.4 & 0 & 0.6 
\end{tabular}
\end{table}
For a general background, the comoving volume element is related to 
$z$ through,
\be
\label{vol}
dV = 4 \pi \left(\frac{c}{H_0}\right) \, r^2 \, 
{\cal F}(\Omega_M,\Omega_{\Lambda},z) \, dz,
\ee
where $H_0$ is the Hubble constant and 
\be
{\cal F}(\Omega_M,\Omega_{\Lambda}, z) \equiv 
\frac{1}{\sqrt{(1+z)^2\,(1+\Omega_M\,z)-z\,(2+z)\,\Omega_
{\Lambda}}}.
\ee
In eq. (\ref{vol}) \op r\cl is the comoving distance,
\begin{eqnarray}
r=\frac{c}{H_0 \sqrt{\Omega_\kappa}}\, S(\sqrt{\Omega_\kappa} \, \int_0^z 
[(1+z')^2 \, (1+ \Omega_M \,z') - \\  
z'\,(2+z')\,\Omega_{\Lambda}]^{-1/2} \, dz')\nonumber
\label{comdist}
\end{eqnarray}
and the function \op S \cl has the following expression (Kim et al. 1997)
\beq
\Omega_M +\Omega_{\Lambda}>1 &S(x)=\sin(x),
& \Omega_\kappa=1-\Omega_M +\Omega_{\Lambda} ,\\
\nn
\Omega_M +\Omega_{\Lambda}<1 
&S(x)=\sinh(x) &\Omega_\kappa = \Omega_M +\Omega_{\Lambda}-1 ,\\
\nn
\Omega_M +\Omega_{\Lambda} =1 &S(x)=x &\Omega_\kappa=1 .
\eeq
Consistently, the SFR history shown in Figure 1 needs to be
properly rescaled in order to account for the cosmological
background dependence of both the luminosity distance and 
the comoving volume within each redshift bin. 

By using equation (\ref{vol}) we have evaluated the function
\op R_{NS}(z) \cl  defined in eq. (\ref{rate}) for the three 
cosmological models, and the results are shown in Figure 2. 

\begin{figure}
\begin{center}
\leavevmode
\centerline{\psfig{figure=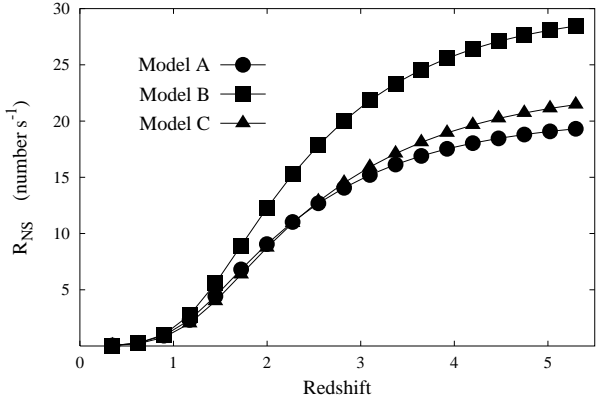,angle=270,width=8cm}}
\caption{Evolution of the rate of neutron star formation
for three different cosmological scenarios.}
\end{center}
\label{fig.rate}
\end{figure}

It should be mentioned that the evaluation of the rate of neutron star 
formation, based on the UV-optically estimated star formation rate evolution
$\dot{\rho}_*(z)$, and on an assumed universal IMF, is largely independent 
of the specific IMF adopted (Madau 1998). In fact, the stars which mainly 
contribute to the UV-optical luminosity are the most massive ones which, 
in turn, are those that at the end of their evolution give origin to 
core-collapse supernovae.

The total number of supernova explosions per unit time 
leaving behind a neutron star is given in Table 1 for the 
considered cosmological models, and for two values of the upper mass cutoff,
 $M=25 \msun$ and  $M=20 \msun$.
As a consequece of the largest number of stars having masses
$< 25 \msun$, the total rate of core-collapses to neutron stars
is a factor $\sim 4$ larger than that of core-collapses to
black holes considered in Ferrari, Matarrese \& Schneider (1998). 
\begin{table}
\centering
\caption{Values of the total rate of neutron stars in different
cosmological scenarios and for two values of the upper progenitor
mass cutoff ($20 \msun$, $25 \msun$).} 
\begin{tabular}{@{}lllllllll@{}}
\multicolumn{3}{c}{Model}& \multicolumn{3}{c}{$M_{up}/\msun$} 
& \multicolumn{3}{c}{$R_{NS}/(events per s)$} \\
 &   & & &  & & &    & \\   
 & A & & &25& & &19.5&   \\
 & A & & &20& & & 17.6&  \\
 & B & & &25& & & 28.7&  \\
 & B & & &20& & &25.9&   \\
 & C & & &25& & &21.7&  \\
 & C & & &20& & &19.6&  
\end{tabular}
\end{table}

\section{Gravitational Emission From Rapidly Rotating Neutron Stars}

In this section we shall briefly describe how the timescale 
of the spindown phase and the energy spectrum
of the gravitational radiation emitted during the r-mode instability
have been estimated (Lindblom, Owen \& Morsink 1998; Owen et al. 1998;
Andersson, Kokkotas \& Schutz 1998; Andersson, Kokkotas, Stergioulas 1998). 
These quantities will be  used to compute, 
respectively, the duty cycle and the spectral energy density
of the stochastic background.

The excitation of an r-mode starts as a small perturbation of
the velocity field.
As the amplitude of the mode grows, non-linear hydrodynamic 
effects become important and eventually dominate 
the dynamics. Under the assumption that the star is 
uniformly rotating  with angular velocity $\Omega$, 
and by parametrizing the r-mode  by its 
amplitude $\alpha$, the star can be considered as a system
having only two degrees of freedom, $\Omega$ and $\alpha$.   
In this simple model, the total angular momentum
for the $\ell=2$ r-mode  can be shown to be
\be
J(\Omega,\alpha) = \bigl(\tilde{I}-\frac{3}{2}\,\tilde{J} 
\alpha^2\bigr)\,\Omega\,M\,R^2 
\;,
\ee
where 
\be
\tilde{I}=\frac{8\,\pi}{3 \,M\,R^2}\,
\int_0^R\!\rho\,r^4\,dr  \quad\hbox{and}\quad
\tilde{J}=
\frac{1}{M\,R^2}\,\int_0^R \, \rho r^6 dr \;,
\ee
and the rotational energy is
\be
\label{erot}
E \approx J(\Omega,\alpha) \Omega.
\ee

Following Owen et al. (1998)  we shall consider
a Newtonian  model of neutron star composed of a fluid with a
polytropic equation of state, $p= k\,\rho^2$, 
with $ k$ chosen so that \op M= 1.4 \msun\cl 
and  \op R=12.53$ km.  For this model,
$\tilde{I}=0.261$ and $\tilde{J}=1.635\times 10^{-2}$. 
It should be noted that $\tilde{I}$ and $\tilde{J}$ depend only on the
chosen polytropic index.

The evolution of the angular momentum of the star is determined
by the emission of gravitational waves, which couple to the r-modes
through the current multipoles, primarily that with $l=m=2$.
For this mode, the frequency of the
emitted gravitational radiation is $\nu=(2/3\pi)\,\Omega$.
The evolution of $\Omega$ and $\alpha$ during the phase
in which $\alpha$ is small, can be determined 
from the standard multipole expression for angular momentum loss, and
from the energy loss due to the gravitational emission and to the  dissipative
effects induced by the bulk and shear viscosity.
In this phase $\Omega$ is nearly constant and
$\alpha$ grows exponentially. After a short time, that Owen et al. estimate
to be of order \op 500$ s, the
amplitude of the mode becomes close to unity and non-linear 
effects saturate and halt a further growth of the mode. This phase
lasts for approximately \op 1$ yr, during which the star loses
angular momentum radiating approximately $ 2/3$ of its
initial rotational energy in gravitational waves,
up to a point where the angular velocity
and the temperature are so low that the r-mode ceases to be unstable.
Viscous forces and gravitational radiation damp out the energy left
in the mode, and the star slowly reaches its final equilibrium configuration.

Since gravitational radiation is emitted over such a long time interval,
the background contributed by neutron stars is continuous.
In fact, the total duty cycle for this process is
\be
D=\int_0^\infty \!\! dR_{NS}(z) \, \overline{\tau}_{NS}
(1+z) \sim 10^{9}.
\ee
It has been suggested that if the proton fraction in the star
is sufficiently large to make the direct URCA process possible,
the star would cool in a much shorter time,
$\overline{\tau}_{NS}\approx 20$ sec.
However, if this were the case,
the duty cycle would still be $\gg 1$ ($D \approx 10^{2}$) and the
resulting background would still be continuous.

It is known that the angular velocity of a rotating star cannot exceed 
the Kepler frequency \op \Omega_K \cl
at which mass shedding at the stellar equator makes the star unstable.
A reasonable approximation for this limiting frequency is
\be
\label{omegak}
\Omega_K \approx C\, \sqrt{\frac{M}{\msun} \, \left(\frac{10\, km}{R}\right)^3}
\;,
\ee
where $C \approx 7.8 \times 10^3 \, \mbox{sec}^{-1}$ (Friedmann, Ipser \& 
Parker 1989), and \op M\cl and \op R\cl refer to the mass and 
radius of the corresponding non-rotating star. 
Following Owen et al., we shall assume that 
neutron stars are born with angular velocity close to the maximum value
$\Omega_K $, and have a rotational  kinetic energy  
\be
E_K \approx J(\Omega_K;\alpha)\,\Omega_K = 
\bigl(\tilde{I}-\frac{3}{2}\,\tilde{J} \alpha^2\bigr)\,M\,R^2\, \Omega_K^2.
\ee
In the non-linear saturation phase, where
most of the gravitational radiation is emitted, $\alpha \approx 1$.
Since in this simple model  approximately 
$2/3$ of the stars rotational energy is  radiated in gravitational waves, 
the expression of the  energy spectrum  can be approximated as follows
\be
\label{degw}
\frac{dE_{GW}}{d\nu} \approx \frac{4}{3} \, E_K\, \frac{\nu}{\nu_{max}^2}.
\ee
where
\be
\label{numax}
\nu_{max} = (2/3\pi) \Omega_K.
\ee
The r-mode instability ceases to be effective for some value
of the rotational frequency \op\Omega_c,\cl which corresponds 
to the final spin period, and which is associated to
a gravitational emission frequency
\be
\label{numin}
\nu_{min}=(2/3\pi)\Omega_c.
\ee 
$\Omega_c\cl can be determined by solving the equation 
$1/\tau(\Omega_c)=0$, where
$\tau$ is the total dissipation timescale which can be decomposed as a sum 
of the damping times associated to the gravitational emission,  to the
shear and to the bulk viscosity. 
$\tau(\Omega_c)\cl is clearly  a function of the temperature of the star,
and it has been shown that the r-mode instability operates only in hot
neutron stars ($T>10^9$ K). Below $10^9$ K, superfluidity
and other non-perfect fluid effects become important and the
damping  due to  viscosity dominates with respect to the destabilizing
effect of the gravitational radiation.
To be consistent with the analysis by Owen et al. (1998), we shall
adopt a value of $\Omega_c \approx 566$ Hz for the final angular velocity 
which corresponds to a final spin period of $\approx 11$ ms and to
$\nu_{min} \approx 120$ Hz.

The qualitative picture that arises from this simple model is believed to
be sufficiently reliable, even though various uncertainties and approximations
might affect the quantitative results for the initial rotation of the star
after collapse, for the spindown timescales 
as well as for the final rotation period 
(Andersson, Kokkotas \& Schutz 1998).
Thus, it will be interesting to evaluate the effects that different values of 
$\nu_{max}$, $E_K$ and $\nu_{min}$ have on our main conclusions.

\section{Spectral Properties of The Stochastic Background}

In order to evaluate the spectral energy density of the stochastic background
produced by the spindown radiation from  newly born
neutron stars, we need to know how the flux emitted by a single source at 
redshift $z$ would be observed today, and convolve it with the differential
rate of the sources $dR_{NS}(z)$ as in Ferrari, Matarrese \& Schneider (1998).

The average energy flux per unit frequency 
emitted by a single neutron star at cosmological
redshift $z$ is
\be
\label{f1}
f(\nu^{obs})=
\frac{1}{4\pi d_L(z)^2} \left(\frac{d E_{GW}}{d\nu}\right)^{obs},
\ee
where \op \nu^{obs}\cl is the frequency at which the radiation would 
presently be observed. The energy flux \op f(\nu^{obs})\cl can be evaluated 
by making use of eq. (\ref{degw}) suitably redshifted as follows
\be
\label{fobs}
f(\nu^{obs})=
\frac{E_K}{3 \pi}\,\frac{\nu^{obs}}{(\nu^{obs}_{max})^2}\, \frac{1}{(1+z)\, 
d_L(z)^2} \;,
\ee
where
\be
\frac{\nu^{em}_{min}}{1+z} \leq
\nu^{obs} \leq \frac{\nu^{em}_{max}}{1+z},
\ee
and $d_L(z) = (1+z) \, r$ is the luminosity distance, which depends
on the cosmological model as indicated in eq. (\ref{comdist}).
Thus, the spectral energy density of the produced background is
\be
\frac{dE_{GW}}{dt dS d\nu} = \int_0^{\infty} \! f(\nu)\, dR_{NS}(z),
\ee
where the differential rate of sources is given by
\be
dR_{NS}(z)= \dot{\rho}_*(z)\, \frac{dV}{dz} \,\left[ 
\int_{8\msun}^{25\msun}\! \Phi(M) \, dM\right] dz .
\ee
[see eq. (\ref{rate})].
From the spectral energy density  the corresponding 
values of the closure energy density of gravitational waves
\be
\Omega_{GW}(a,\nu_{obs})= \frac{\nu_{obs}}{c^3 \rho_{cr}}\, \frac{dE}{dt dS
d\nu},
\ee
where \op \rho_{cr}= 3H_0^2\,/\,8\pi ~G,\cl
and of the spectral strain amplitude 
\be
\label{strain}
\sqrt{S_h(\nu_{obs})}=\left(\frac{2 G}{\pi c^3}
\frac{1}{\nu_{obs}^2}\right)^{1/2}\left(\frac{dE}{dt dS d\nu}\right)^{1/2}
\ee
can be derived.

In Figures 3 and 4 we plot the spectral energy density and the
strain amplitude  for the flat background
with zero cosmological constant corresponding to model A (see Table 1).
The effect of a varying cosmological background is negligible on both 
the spectrum and the strain amplitude. In fact, the amplification
of the rate at high redshifts shown in Figure 2 for the models B and C,
is mostly suppressed by
the inverse squared luminosity distance dependence of the single
source spectrum [see eq. (\ref{fobs})] for the same models. 
Conversely,  the closure 
density exhibits a more evident dependence on the Hubble constant,
as  shown in Figure 5, where we
have considered only model A and B because the dependence on the other
cosmological parameters (namely $\Omega_M$ and $\Omega_{\Lambda}$) is
negligible.

\begin{figure}
\begin{center}
\leavevmode
\centerline{\psfig{figure=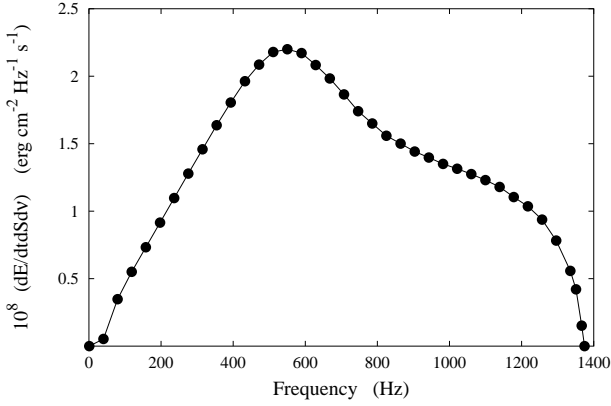,angle=270,width=8cm}}
\caption{The spectral energy density  $(dE/dSdtd\nu)$
is plotted  as a function of the observational frequency for the cosmological
background corresponding to model A (see Table 1).}
\end{center}
\end{figure}
\begin{figure}
\begin{center}
\leavevmode
\centerline{\psfig{figure=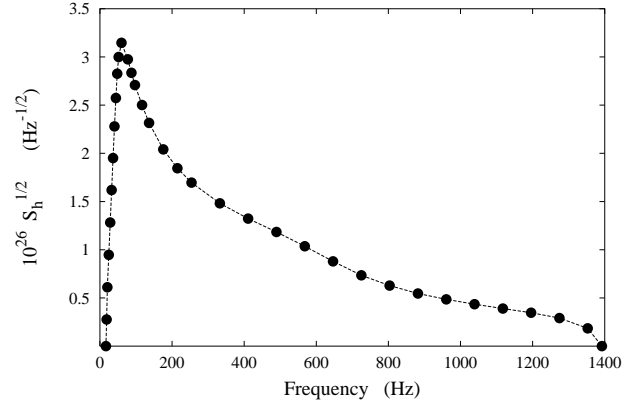,angle=270,width=8cm}}
\caption{The spectral strain amplitude $S_h^{1/2}$ corresponding to the
spectral energy density given in Figure 3 is plotted  as a function of the
observational frequency.}
\end{center}
\end{figure}
\begin{figure}
\begin{center}
\leavevmode
\centerline{\psfig{figure=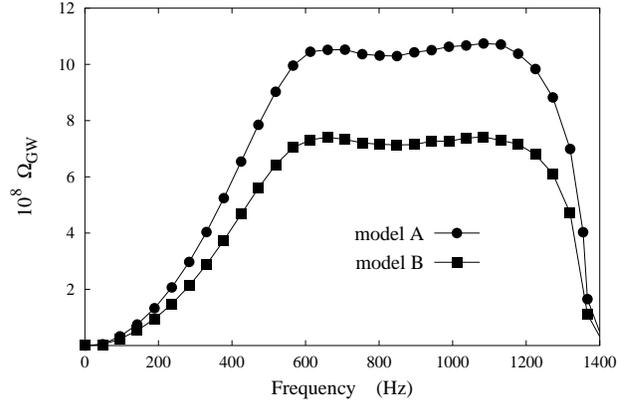,angle=270,width=8cm}}
\caption{The function $\Omega_{GW}$
corresponding to the spectral energy density plotted in Figure 3. 
Since the main effect on the gravitational background is due to the 
change in the Hubble constant, we have plotted only two curves corresponding to
an $h=0.5$ flat model without cosmological constant and to an $h=0.6$
flat model with a cosmological constant (see Table 1).}
\end{center}
\end{figure}

The estimate of the intensity of the stochastic background also depends
on the chosen values of \op\nu_{min}\cl and \op\nu_{max}.\cl 
We have repeated our calculations for two values  of $ \nu_{min}, $
chosen so as to cover the range of pulsar rotational periods, 
$P_{in}$, inferred from observations
\be
\nu_{min}=\frac{4}{3} P_{in}^{-1} \;,
\ee
with $P_{in}$ between 9 and 22.8 ms, where the upper limit is the
maximum rotational period at which the r-modes are unstable
(Andersson, Kokkotas \& Schutz 1998). 
Since the strain amplitude is sensitive mainly to the low frequency
region of the spectrum,  due to the factor $ \nu_{obs}^{-2} $
in eq. (\ref{strain}), a decrease of \op\nu_{min}\cl determines an
increase of the peak amplitude and shifts its location
towards lower frequencies, as shown in Figure 6.

The value of $\nu_{max}$, depends on $ \Omega_K $ which, in turn,
depends on the mass and radius of the star,
as shown in eqs. (\ref{numax}) and (\ref{omegak}). The plots  shown
in Figures 3, 4 and 5 were obtained by assuming $M=1.4 \msun$ and
$R=12.53$ km.  In reality, the possible radii for
rotating neutron stars range from $10$ to $15$ km, therefore, since the
dependence of $ \Omega_K $ on the mass is less significant than that on
the radius, we have computed the strain amplitude and the closure density
by considering $M=1.4 \msun$, and changing the radius to  $R=10$ km and 
to $R=15$ km. 
The effect of reducing  $\nu_{max}$, i.e. of increasing  the radius,
is that of  narrowing the interval of
frequencies over which each source contributes. Since the energy
flux of a single event  does not depend on
$ \Omega_K,$ [see eq. (\ref{fobs})],
 this will result in an increase of the amplitude of the peak
of the  spectral energy density.
This effect is amplified in the strain amplitude, that
is more sensitive to low frequencies.
However, it should be stressed that the increase is  smaller than a factor
2. Conversely, the closure density is more sensitive to the 
high  frequency contributions, and
therefore the effect is the opposite.
The results are shown in  Figures 7 and 8.

\begin{figure}
\begin{center}
\leavevmode
\centerline{\psfig{figure=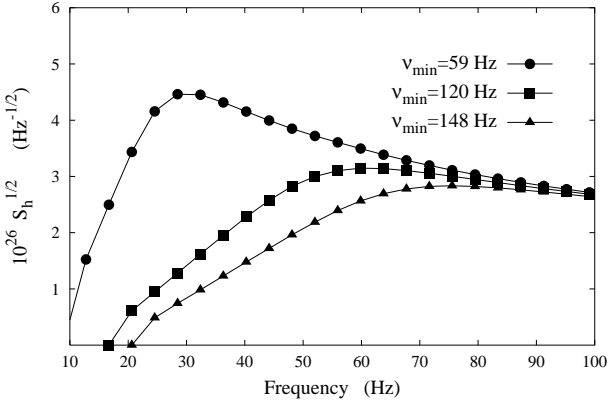,angle=270,width=8cm}}
\caption{The spectral strain amplitude $S_h^{1/2}$ corresponding to a
flat model is plotted  as a function of the 
observational frequency for three different values of $ \nu_{min} $
(see text).}
\end{center}
\end{figure}
\begin{figure}
\begin{center}
\leavevmode
\centerline{\psfig{figure=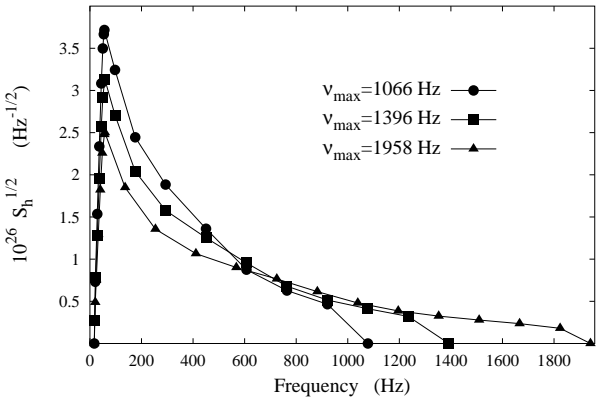,angle=270,width=8cm}}
\caption{The spectral strain amplitude $S_h^{1/2}$ corresponding to a
flat model is plotted  as a function of the 
observational frequency for three different values of the 
radius and, thus, of the initial rotational period and of the total 
rotational energy (see text).}
\end{center}
\end{figure}
\begin{figure}
\begin{center}
\leavevmode
\centerline{\psfig{figure=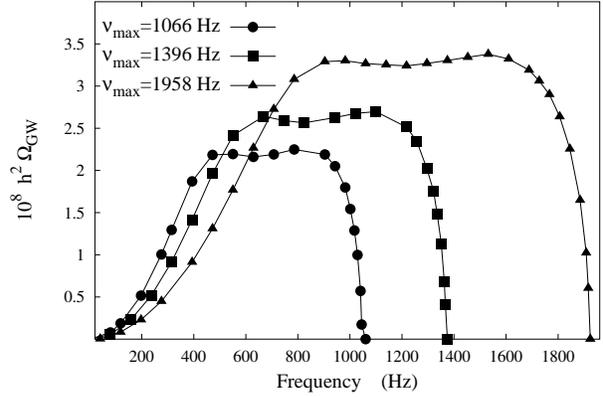,angle=270,width=8cm}}
\caption{The function $\Omega_{GW}$
corresponding to the spectral strain amplitude plotted in Figure 7.}
\end{center}
\end{figure}

\section{Detectability} 
The value of the optimal signal-to-noise ratio (SNR) for L-shaped 
interferometers can be computed according to the formula (Allen 1996)
\be
\label{SNRint}
SNR^2=\left[\left(\frac{9 H_0^4}{50\pi^4} \right) T
\int_0^\infty  d\nu \frac{\gamma^2(\nu)\Omega^2_{GW}(\nu) }
{\nu^6 S_h^{(1)}(\nu) S_h^{(2)}(\nu)}
\right]
\ee
where \op S_h^{(i)}\cl are the spectral noise densities of two detectors
operating in coincidence, and \op\gamma(\nu)\cl is the overlap function
which takes into account the difference in their location and orientation.
We shall calculate SNR for several pairs of detectors, and
for $ \Omega_{GW}$ computed for a  flat  cosmology,
$\nu_{min}=120 $ Hz, and  $ \nu_{max}= 1396$ Hz.
We shall assume an integration time of $T=1$ yr.

For interferometric antennas, 
we have computed \op S_h^{(i)}\cl  from the expected sensitivity curves
given for VIRGO (Beccaria et al. 1996) and for ADVANCED LIGO (Flanagan \& 
Hughes 1997). The sensitivity curve for GEO600 has been extracted
from the  Web site http://www.geo600.uni-hannover.de.

For VIRGO and GEO600 we find $SNR=3.7\times 10^{-3}$,
assuming optimal orientation, and $SNR=5.0\times 10^{-3}$
if GEO600 is narrowbanded around $100$ Hz.
For two ADVANCED LIGO detectors $SNR=0.84$.
Since the correlation of two interferometers at a distance $d$ 
starts to be very poor around a frequency $\approx (70 \, \mbox{Hz})
(3000\, \mbox{km}/d)$ (Allen \& Romano 1998), two nearby detectors 
respond better to higher frequencies, where our predicted signal has a larger
amplitude (see Figure 8).
Thus, the signal to noise ratio rises to $8.5$ if we consider two
nearby ADVANCED LIGO (or VIRGO) interferometers at a distance 
of $\approx 300$ km.

We have similarly investigated how interferometric and resonant 
detectors operating in coincidence might respond to our predicted 
background. 
The correlation of VIRGO and NAUTILUS, situated $265$ km apart, gives 
$SNR=3.0\times 10^{-4}$. Assuming that a similar 
correlation is done with a truncated icosahedral antenna
(TIGA) (Merkowitz \& Johnson 1995) and an ADVANCED LIGO interferometer
located at the same site, the resulting $SNR$ is $\approx 1.3\times 10^{-3}$,
whereas it is $SNR~\approx~1.7\times 10^{-3}$ if the interferometer
is  ADVANCED VIRGO.

Thus, the detection of this background requires two nearby interferometers
with `advanced' sensitivities, as similarly noted by Owen et al. (1998).
 
\section{Conclusions} 

In this paper we have evaluated  the gravitational 
background produced by a cosmological population of  
hot, young and rapidly rotating neutron stars, emitting gravitational
radiation during the spin-down phase associated to the r-mode instability.
The dependence of this background on the
cosmological model and on the parameters that enter into the 
determination of the spectrum of each single event, have been discussed.
Our predictions for the spectrum, spectral strain amplitude and closure 
density, appear to be rather insensitive
to the assumed cosmological parameters. This is because the increase of
the neutron star birth rate we find in an open or a flat
low-density cosmology is compensated by the increase of the
luminosity distance as a function of redshift,
which suppresses the contribution of the farthest sources.

For the same reason, the uncertainties that still trouble the high redshift
tail of the SFR density evolution due to dust obscuration, are not
severe for our signal, as this is mainly contributed by low-to-intermediate
redshift sources. Moreover, if dust obscuration produces an upward correction
of the SFR density at intermediate redshifts (i.e. around the location of
the maximum) then the amplitude of $(S_h)^{1/2}$ and of $\Omega_{GW}$ 
will be similarly amplified at $\approx 30~ Hz $ and $\approx 600$ Hz 
respectively.

We find that, within a reasonable range of values of the main parameters 
which characterize the energy spectrum of a single source (namely $\nu_{min}$,
$\nu_{max}$ and $E_{k}$), the spectral strain amplitude  has a
maximum ranging within $\approx (2-4)\times 10^{-26} \, 
\mbox{Hz}^{-1/2}$ at frequencies $\approx (30-60)$ Hz,
whereas the closure density, $h^2 \, \Omega_{GW}$, is shown to have
a long plateau extending from $\approx 300$ Hz up to $\approx 1700$ Hz with
an amplitude of $\approx (2.2-3.3)\times 10^{-8}$.

It is interesting to compare our results with those of a similar study 
done by Owen et al. (1998), who assume a constant 
comoving number density of neutron star births in the range $0<z<4$, and
zero elsewhere.

\begin{figure}
\begin{center}
\leavevmode
\centerline{\psfig{figure=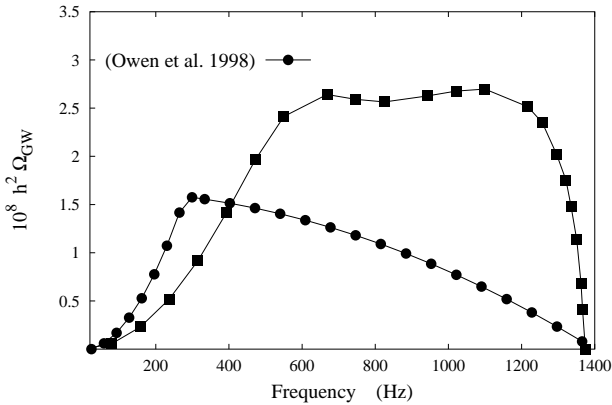,angle=270,width=8cm}}
\caption{A comparison with the closure density, $h^2 \, \Omega_{GW}$,
predicted by Owen et al. (1998) is plotted as a function of the
observational frequency for the cosmological background corresponding to
model A.}
\end{center}
\end{figure}
In Figure 9 we show the closure density predicted by the present analysis
and that obtained by Owen et al.
The two curves exhibit quite a different high frequency behaviour.
Since the cosmological expansion shifts the frequencies towards 
smaller values, the high frequency tail corresponds to closer sources,
thus the difference between the two curves can be traced back 
to  the SFR density evolution between $z=0$ and $z \approx 1-2$. 
The peak at $\approx 600$ Hz exhibited by our curve,
corresponds to $z \approx 1.3$, where
we expect to have the maximum number of sources. Conversely, the maximum
amplitude predicted by Owen et al. is located around $280$ Hz, and 
corresponds to the energy emitted at maximum spin rate by a source at
$z\sim 4$. Thus, their signal appears to be mostly contributed by distant
sources near their maximum rotational velocity, and this is due 
to the fact  that they do not consider the luminosity distance damping,
which is particularly effective for high redshift sources.
It is important to stress that in the region where cross-correlation between
two LIGO interferometers can be accomplished, i.e. for $\nu < 50-60$ Hz, the
signal is entirely emitted at $z>1$, where it is important to embed the
gravitational sources in the correct cosmological environment.

Both in our analysis and in the study of Owen et al.,
all neutron stars have been assumed to be initially rotating with
an angular velocity close to the Keplerian
value, while there might be a fraction of stars born with lower 
initial spin rate. 
In particular, no spindown gravitational radiation would  be
emitted if the initial angular velocity of the star is too small for the
instability to set in Spruit \& Phinney (1998).
However, according to Andersson, Kokkotas \& Schutz (1998) this 
is unlikely to be the generic situation, because
the angular momentum of the degenerate core required
to make the star rotating with velocity $\Omega_K$ is rather small.
Moreover, there is clear evidence that some pulsars were born spinning
rapidly: as an example,
the observed spin period for the young X-ray pulsar in the
supernova remnant N157B is 16 ms, and its inferred initial spin
period is estimated to be $\approx 6-9$ ms (Marshall et al. 1998).

\section*{Acknowledgments}

We greatly acknowledge Pia Astone, Piero Madau, Cedric Lacey
and Silvia Zane for useful discussions.

\label{lastpage}

\end{document}